# Mixsourcing: Exploring Bounded Creativity as a Form of Crowdsourcing

Sarah Hallacher[1,2]   Jenny Rodenhouse[1]   Andrés Monroy-Hernández[1]

[1] Microsoft Research                    [2] NYU ITP

## ABSTRACT
In this paper, we introduce the concept of mixsourcing as a modality of crowdsourcing focused on using remixing as a framework to get people to perform creative tasks. We implement this idea through two design exercises that helped us identify the promises and challenges of this peer-production modality.

## Author Keywords
creativity; art; remixing; crowdsourcing; mixsourcing;

## ACM Classification Keywords
J.5 [Arts and Humanities]: Media Arts; H 5.3. [Information Interfaces and Presentation]: Group and Organization Interfaces—Computer supported cooperative work.

## General Terms
Design

## INTRODUCTION
Crowdsourcing creative work, especially without financial incentives, has remained an elusive goal for designers of peer-production system. For example, one of the most prominent crowdsourcing platforms, Amazon Mechanical Turk, has largely been used for machine-like tasks that often lack creativity and expressiveness. Moreover, even when artists make use of crowdsourcing platforms, crowd-workers tend to be used as mere instruments to achieve a larger vision, rather than recognized as unique individuals. For example, participants of the successful Johnny Cash Project [3] or the Sheep Market [4], are simply instructed to draw image frames that are later aggregated into an undifferentiated mass.

In contrast, *remixing*, such as the creation of video mashups or humorous image macros [5,7], has thrived as a form of unbounded creative peer-production. However, remixing is often unconstrained and ad-hoc, lacking some of the uniformity and consistency of more mechanical crowdsourcing techniques.

In this paper, we ask whether remixing can be reused as a mechanism to crowdsource creative tasks, which we refer to as *mixsourcing*. We explore this via a two-phase design experiment with a group of art students, and by implementing a prototype based on the lessons learned. In the next section, we describe the design experiment where we investigated the conditions under which "bounded creativity" through mixsourcing elicits voluntary participation, and the type of creative output it can inspire. We then present and evaluate a prototype that embodied these ideas. We conclude with a set of implications for design. Our paper is accompanied by a video to help the reader understand a variety of usage scenarios for our prototype.

## DESIGN EXPERIMENT
Inspired by similar exercises in improv theater [6] and the Surrealist "exquisite corpse" technique [1], the first phase of exploring mixsourcing consisted in giving a group of 22 art students a personalized creative task seeded by a hand-made piece of art. The seed content used in this first exercise was an image, hand-drawn by one of the researchers, showing a "moonicorn": a unicorn with a moon as a head (See Figure 1).

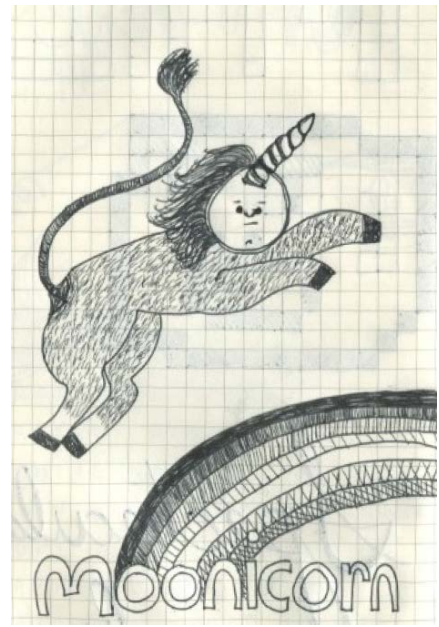

**Figure 1: Hand-drown image provided by researcher.**

The image was accompanied by a message that appealed to the social relationship between group members and with one of the researchers. Each person was offered a task: an invitation to turn the moonicorn into something else, based on our knowledge of their talents. The message read as follows:

> "Hello my dear friends! [...] I'm writing to see if you can help me with my summer project [...] I gave you all top secret assignments below. If you can help, it would mean so so so so so much to me. You don't have to spend a ton of time on it. And I'll throw a boozy thanks-you party next week when I'm in town. I couldn't ask for a better lump of friends. Much love from the west coast. This is my moonicorn (see Figure 1)."

The message then went on at length to describe each person's assignment. For example, one person named "Joe" was asked to "create a moonicorn cocktail recipe," while "Kate" was asked to collaborate with "Matt" to "create an iPhone video documenting the rare, nut-eating moonicorn, played by @Matt."

The second phase of the design experiment consisted in an attempt to scale up the first phase. We used a school mailing list and a Facebook Group to send the same message and seed content, without the personalized tasks. The message was an open plea to participate in the remixing of the moonicorn or even just to express what they would turn the moonicorn into.

### Results
The first phase of the experiment was quite successful. More than half of the art students (12) actually completed their remix, and most of them spent significant time on it. The majority were as eager to narrate their creative process as to share their finished work. For example, "Joe"[1] not only wrote his "moonicorn cocktail recipe," he also prepared one, took a picture of it, and emailed the make-believe recipe to accompany the beverage:

> "The Sanguine Moonicorn. 6oz Fresh Moonicorn Blood. 1oz Pure Moonicorn Tears [hold for virgins] Topped with Moonicorn Sweetbreads, Moonicorn Gonad, Moon Cheese, and an olive. Cleansed by fire and served over ice, with a Moonicorn Jerky Moon Dagger."

Likewise, "Kate" and "Matt" completed their collaborative task and produced a short video. Other remixes we received included a fiction article, a felt toy, and a music mix (see Figure 2 for a collage of all remixes received).

The second phase was not as successful however; of the more than 200 people we reached with the mailing list and the Facebook Group only 11 responded with a very short

---

[1] All names are fictional for blind review and privacy of the participants.

description of what they would do with the moonicorn, without actually creating a remix of it.

We followed up with participants of the first phase of the experiment to ask more about their experience. We asked what motivated them to participate, and most reported participating because of their relationship with the researcher "Jane" and the novelty of the exercise. For example, "Andrea" explained: "very much respect and admire Jane's work-- and was flattered to get individual direction from her. Her direction definitely stemmed from an understanding of my skillset-- though I would also have enjoyed the challenge to step outside of my skillset as well. The specific direction though was key-- having the impetus to create that's outside of work or personal persuasions was incredibly refreshing."

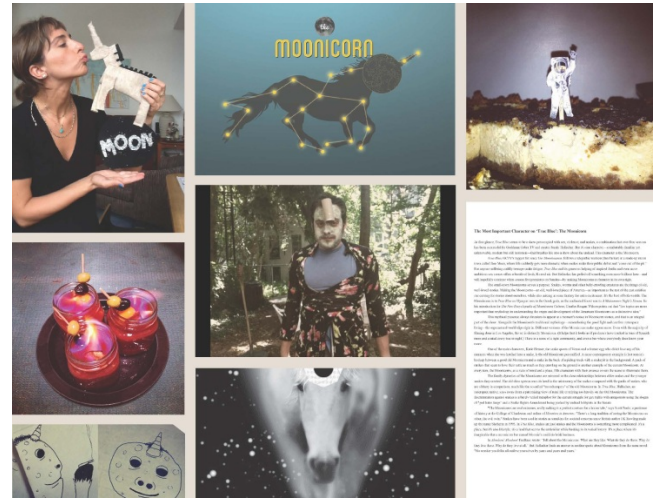

**Figure 2: Remixes of the first exercise**

### Discussion
While these were not formal, controlled experiments, an analysis of both phases of the exercise remains useful. We conjecture that there were three key attributes that contributed to the success of the first phase and the failure of the second:

1. Pre-existing personal relationships. Unlike the second phase, the first built on strong ties between the participants and the researcher.
2. Hand-crafted, personalized tasks directed at specific individuals, compared to the diffusion of responsibilities, according to well-described precepts in the social psychology literature [2].
3. Specificity of tasks. The message to the broader group was too open, putting the burden of choosing on the remixers.

### PROTOTYPE: "TURN THIS INTO THAT"
Using the insights gained from the previous exercise, we envisioned a mixsourcing platform to enable people to create and participate in remixing exchanges like the moonicorn one. We called the platform "Turn This Into That," as it describes the system's premise.

The system allows people to create and participate in challenges. An example of a challenge could be something like "Turn this scene out of the movie 'Eternal Sunshine of a Spotless Mind' into a photo of you." This structure is detailed enough to guide the contribution, while leaving the choice of what to do and how to do it up to the remixer responding to the challenge.

To convey the spirit of the social relationships that we thought were instrumental to the success of the first exercise, we built the system on a postcard metaphor. People can send postcards challenging one another to "turn something into something else," and the responses themselves are also in the form of postcards (see Figure 3, Figure 4, and Figure 5). All postcards begin with the phrase "Dear everyone," followed by the optional name of a particular person whom the challenger or the remixer might want to reach out to.

Furthermore, given the interest people showed in talking about their creative process, the submissions provide space and encouragement for people to address the why's and the how's of their challenges and their remixes.

Practically speaking, this platform was designed as a simple mechanism to provide creative sparks, playfulness, and interactions among people without actually having to deal with the complexities of creating tools or even repositories for the content itself. The website's home page[2] features a list of available challenges that invite people to remix (see Figure 6). Also system relies completely on the social media ecosystem for all that, and *Turn This Into That* simply facilitates the linkages between people through creative triggers in the form of postcards.

A couple of days after going live, we asked a couple of people to try out the website and provide feedback. In the next section we share what we learned.

### Results

While still early in its existence, the website already has a half a dozen remix challenges, most of which have elicited at least one response. When asked whether they would continue using the website people mentioned that they would because they found it to be an enjoyable activity and put it in the context of other similar websites they used. For example, "Brad" mentioned, "it could be pretty fun. It's definitely a cross between Facebook and Pintrest, where you get to explore random interesting stuff, but also see people you know flex their creative muscles." Similarly, "Jake" mentioned, "I could see myself checking back to see what people have done or post my own challenge like for music I've written, I'd have people remix it. I don't know if I'd use it for work but I know I'd look at it while at work to give me ideas."

### CONCLUSIONS

There are an increasing number of websites that "mashup" web platforms by connecting them one with the other; however, we believe there is still a great deal of unmined potential in using virtual platforms to connect people for creative interaction and generativity. In this paper we have proposed one such service and built a prototype based on what we found to be a successful low-tech experiment.

We have presented both a set of design exercises and a system prototype to explore a new soft-touch crowdsourcing modality. We believe this is a rich area for future exploration in catalyzing and supporting creative collaboration and large-scale artistic interactions building on top of the existing social media infrastructure.

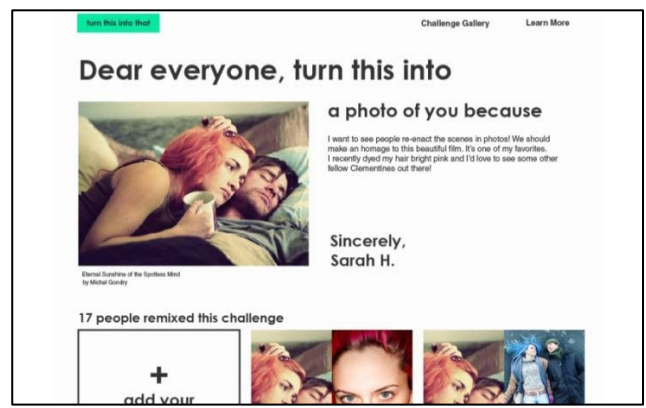

**Figure 3: Form to submit remixing challenge.**

**Figure 4: Challenge**

---

[2] Accessible at http://turnthisinto.com

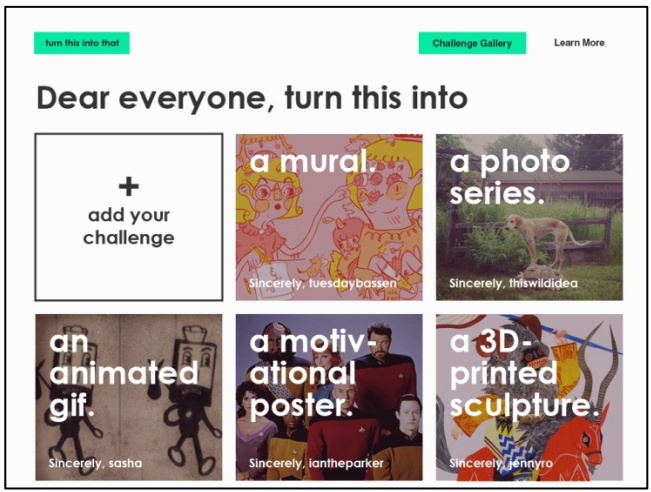

**Figure 5: Form to submit remix**

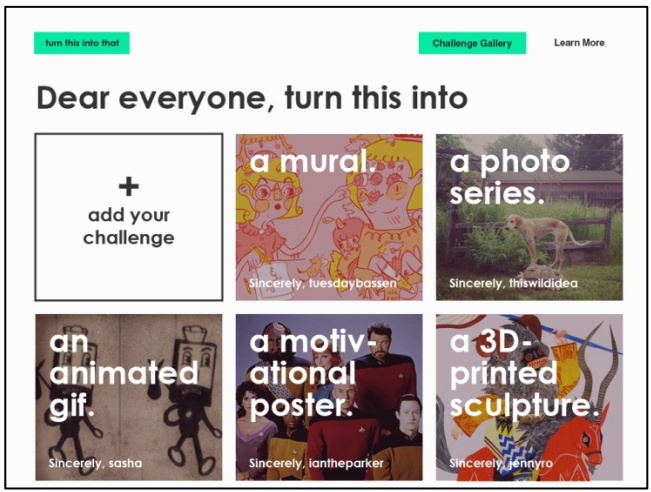

**Figure 6: Website's Home page.**